\pdfoutput=1

\documentclass[12pt,a4paper]{article}

\usepackage{ifthen} 
\newboolean{pdflatex}
\setboolean{pdflatex}{true} 

\newboolean{articletitles}
\setboolean{articletitles}{true} 

\newboolean{uprightparticles}
\setboolean{uprightparticles}{false} 

\newboolean{inbibliography}
\setboolean{inbibliography}{false} 

\usepackage{microtype}
\usepackage{lineno}  
\usepackage{xspace} 
\usepackage{caption} 

\usepackage{graphicx}  
\usepackage{color}
\usepackage{colortbl}
\graphicspath{{./figs/}} 

\usepackage{amsmath} 
\usepackage{amssymb}
\usepackage{amsfonts}
\usepackage{upgreek} 

\newcommand*\patchAmsMathEnvironmentForLineno[1]{%
\expandafter\let\csname old#1\expandafter\endcsname\csname #1\endcsname
\expandafter\let\csname oldend#1\expandafter\endcsname\csname
end#1\endcsname
 \renewenvironment{#1}%
   {\linenomath\csname old#1\endcsname}%
   {\csname oldend#1\endcsname\endlinenomath}%
}
\newcommand*\patchBothAmsMathEnvironmentsForLineno[1]{%
  \patchAmsMathEnvironmentForLineno{#1}%
  \patchAmsMathEnvironmentForLineno{#1*}%
}
\AtBeginDocument{%
\patchBothAmsMathEnvironmentsForLineno{equation}%
\patchBothAmsMathEnvironmentsForLineno{align}%
\patchBothAmsMathEnvironmentsForLineno{flalign}%
\patchBothAmsMathEnvironmentsForLineno{alignat}%
\patchBothAmsMathEnvironmentsForLineno{gather}%
\patchBothAmsMathEnvironmentsForLineno{multline}%
\patchBothAmsMathEnvironmentsForLineno{eqnarray}%
}

\usepackage{hyperref}    
\usepackage[all]{hypcap} 

\usepackage{xspace} 
\usepackage{upgreek}

\def\lhcb {\mbox{LHCb}\xspace}

\def\MagUp {\mbox{\em Mag\kern -0.05em Up}\xspace}

\ifthenelse{\boolean{uprightparticles}}%
{

 \def\Pmu         {\ensuremath{\upmu}\xspace}

 \def\Ppi         {\ensuremath{\uppi}\xspace}

 \def\Ppsi        {\ensuremath{\uppsi}\xspace}

 \def\PDelta      {\ensuremath{\Delta}\xspace}                 
 \def\PXi      {\ensuremath{\Xi}\xspace}                 
 \def\PLambda      {\ensuremath{\Lambda}\xspace}                 
 \def\PSigma      {\ensuremath{\Sigma}\xspace}                 
 \def\POmega      {\ensuremath{\Omega}\xspace}                 
 \def\PUpsilon      {\ensuremath{\Upsilon}\xspace}                 
 
 \def\PB      {\ensuremath{\mathrm{B}}\xspace}                 
                  
 \def\PD      {\ensuremath{\mathrm{D}}\xspace}

 \def\PJ      {\ensuremath{\mathrm{J}}\xspace}                 
 \def\PK      {\ensuremath{\mathrm{K}}\xspace}

 \def\Pb      {\ensuremath{\mathrm{b}}\xspace}                 
 \def\Pc      {\ensuremath{\mathrm{c}}\xspace}

 \def\Pi      {\ensuremath{\mathrm{i}}\xspace}

 \def\Pp      {\ensuremath{\mathrm{p}}\xspace}

}
{

 \def\Pmu         {\ensuremath{\mu}\xspace}

 \def\Ppi         {\ensuremath{\pi}\xspace}

 \def\Ppsi        {\ensuremath{\psi}\xspace}                 
                  
 \mathchardef\PDelta="7101
 \mathchardef\PXi="7104
 \mathchardef\PLambda="7103
 \mathchardef\PSigma="7106
 \mathchardef\POmega="710A
 \mathchardef\PUpsilon="7107
                  
 \def\PB      {\ensuremath{B}\xspace}                 
                  
 \def\PD      {\ensuremath{D}\xspace}

 \def\PJ      {\ensuremath{J}\xspace}                 
 \def\PK      {\ensuremath{K}\xspace}

 \def\Pb      {\ensuremath{b}\xspace}                 
 \def\Pc      {\ensuremath{c}\xspace}

 \def\Pi      {\ensuremath{i}\xspace}

 \def\Pp      {\ensuremath{p}\xspace}

}

\makeatletter
\ifcase \@ptsize \relax
  \newcommand{\miniscule}{\@setfontsize\miniscule{4}{5}}
\or
  \newcommand{\miniscule}{\@setfontsize\miniscule{5}{6}}
\or
  \newcommand{\miniscule}{\@setfontsize\miniscule{5}{6}}
\fi
\makeatother

\DeclareRobustCommand{\optbar}[1]{\shortstack{{\miniscule (\rule[.5ex]{1.25em}{.18mm})}
  \\ [-.7ex] $#1$}}

\def\mup        {{\ensuremath{\Pmu^+}}\xspace}
\def\mun        {{\ensuremath{\Pmu^-}}\xspace} 

\def\cquark    {{\ensuremath{\Pc}}\xspace}

\def\bquark    {{\ensuremath{\Pb}}\xspace}

\def\pion   {{\ensuremath{\Ppi}}\xspace}

\def\pip    {{\ensuremath{\pion^+}}\xspace}
\def\pim    {{\ensuremath{\pion^-}}\xspace}

\def\kaon    {{\ensuremath{\PK}}\xspace}
  \def\Kbar    {{\kern 0.2em\overline{\kern -0.2em \PK}{}}\xspace}

\def\KorKbar    {\kern 0.18em\optbar{\kern -0.18em K}{}\xspace}

\def\Km      {{\ensuremath{\kaon^-}}\xspace}

\def\Kstarzb {{\ensuremath{\Kbar{}^{*0}}}\xspace}
\def\Kstar   {{\ensuremath{\kaon^*}}\xspace}

  \def\Dbar    {{\kern 0.2em\overline{\kern -0.2em \PD}{}}\xspace}
\def\D       {{\ensuremath{\PD}}\xspace}

\def\DorDbar    {\kern 0.18em\optbar{\kern -0.18em D}{}\xspace}
\def\Dz      {{\ensuremath{\D^0}}\xspace}

\def\Dstarp  {{\ensuremath{\D^{*+}}}\xspace}

\def\Bbar    {{\ensuremath{\kern 0.18em\overline{\kern -0.18em \PB}{}}}\xspace}

\def\BorBbar    {\kern 0.18em\optbar{\kern -0.18em B}{}\xspace}

\def\jpsi     {{\ensuremath{{\PJ\mskip -3mu/\mskip -2mu\Ppsi\mskip 2mu}}}\xspace}

  \def\Y#1S{\ensuremath{\PUpsilon{(#1S)}}\xspace}

\def\proton      {{\ensuremath{\Pp}}\xspace}

\def\Xires       {{\ensuremath{\PXi}}\xspace}

\def\Lz          {{\ensuremath{\PLambda}}\xspace}
\def\Lbar        {{\ensuremath{\kern 0.1em\overline{\kern -0.1em\PLambda}}}\xspace}
\def\LorLbar    {\kern 0.18em\optbar{\kern -0.18em \PLambda}{}\xspace}

\def\Omegares    {{\ensuremath{\POmega}}\xspace}

\def\Lb      {{\ensuremath{\Lz^0_\bquark}}\xspace}

\def\Lc      {{\ensuremath{\Lz^+_\cquark}}\xspace}

\def\Xibz    {{\ensuremath{\Xires^0_\bquark}}\xspace}
\def\Xibm    {{\ensuremath{\Xires^-_\bquark}}\xspace}

\def\Xicz    {{\ensuremath{\Xires^0_\cquark}}\xspace}

\def\Omegac    {{\ensuremath{\Omegares^0_\cquark}}\xspace}
\def\OmegacStar    {{\ensuremath{\Omegares^{*0}_\cquark}}\xspace}

\def\Omegab    {{\ensuremath{\Omegares^-_\bquark}}\xspace}

\def\Xim    {{\ensuremath{\Xires^-}}\xspace}
\def\Xiz    {{\ensuremath{\Xires^0}}\xspace}
\def\Omegam    {{\ensuremath{\Omegares^-}}\xspace}

\def\to                 {\ensuremath{\rightarrow}\xspace}

\def\AT#1     {\ensuremath{A_{\mathrm{T}}^{#1}}\xspace}           

\def\C#1      {\ensuremath{\mathcal{C}_{#1}}\xspace}                       
\def\Cp#1     {\ensuremath{\mathcal{C}_{#1}^{'}}\xspace}                    
\def\Ceff#1   {\ensuremath{\mathcal{C}_{#1}^{\mathrm{(eff)}}}\xspace}        
\def\Cpeff#1  {\ensuremath{\mathcal{C}_{#1}^{'\mathrm{(eff)}}}\xspace}       
\def\Ope#1    {\ensuremath{\mathcal{O}_{#1}}\xspace}                       
\def\Opep#1   {\ensuremath{\mathcal{O}_{#1}^{'}}\xspace}                    

\newcommand{\tev}{\ifthenelse{\boolean{inbibliography}}{\ensuremath{~T\kern -0.05em eV}\xspace}{\ensuremath{\mathrm{\,Te\kern -0.1em V}}}\xspace}
\newcommand{\gev}{\ensuremath{\mathrm{\,Ge\kern -0.1em V}}\xspace}
\newcommand{\mev}{\ensuremath{\mathrm{\,Me\kern -0.1em V}}\xspace}
\newcommand{\kev}{\ensuremath{\mathrm{\,ke\kern -0.1em V}}\xspace}
\newcommand{\ev}{\ensuremath{\mathrm{\,e\kern -0.1em V}}\xspace}
\newcommand{\gevc}{\ensuremath{{\mathrm{\,Ge\kern -0.1em V\!/}c}}\xspace}
\newcommand{\mevc}{\ensuremath{{\mathrm{\,Me\kern -0.1em V\!/}c}}\xspace}
\newcommand{\gevcc}{\ensuremath{{\mathrm{\,Ge\kern -0.1em V\!/}c^2}}\xspace}
\newcommand{\gevgevcccc}{\ensuremath{{\mathrm{\,Ge\kern -0.1em V^2\!/}c^4}}\xspace}
\newcommand{\mevcc}{\ensuremath{{\mathrm{\,Me\kern -0.1em V\!/}c^2}}\xspace}

\def\mum  {\ensuremath{{\,\upmu\mathrm{m}}}\xspace}

\def\invfb   {\ensuremath{\mbox{\,fb}^{-1}}\xspace}

\newcommand{\chisq}{\ensuremath{\chi^2}\xspace}

\newcommand{\chisqip}{\ensuremath{\chi^2_{\text{IP}}}\xspace}

\def\gsim{{~\raise.15em\hbox{$>$}\kern-.85em
          \lower.35em\hbox{$\sim$}~}\xspace}
\def\lsim{{~\raise.15em\hbox{$<$}\kern-.85em
          \lower.35em\hbox{$\sim$}~}\xspace}

\def\pt         {\mbox{$p_{\mathrm{ T}}$}\xspace}

\def\evtgen     {\mbox{\textsc{EvtGen}}\xspace}

\def\geant      {\mbox{\textsc{Geant4}}\xspace}

\def\photos     {\mbox{\textsc{Photos}}\xspace}

\def\pythia     {\mbox{\textsc{Pythia}}\xspace}

\def\tell1  {TELL1\xspace}
\def\ukl1   {UKL1\xspace}

\def\br{{\cal{B}}}

\def\eff{\epsilon}

\def\eff{\epsilon}

\def\rsigma{r_{\sigma}}
\def\fsigma{f_{\sigma}}

\def\tauLb{\tau_{\lower0.85pt\hbox{\scalebox{0.6}{\Lb}}}}
\def\tauXibz{\tau_{\lower0.85pt\hbox{\scalebox{0.6}{\Xibz}}}}
\def\tauXib{\tau_{\lower0.85pt\hbox{\scalebox{0.6}{\Xibm}}}}
\def\tauOmb{\tau_{\lower0.85pt\hbox{\scalebox{0.6}{\Omegab}}}}

\def\mXib{m_{\lower0.85pt\hbox{\scalebox{0.6}{\Xibm}}}}
\def\mOmb{m_{\lower0.85pt\hbox{\scalebox{0.6}{\Omegab}}}}

\usepackage{cite} 
\usepackage{mciteplus}

\usepackage{longtable} 

\begin{document}

\renewcommand{\thefootnote}{\fnsymbol{footnote}}
\setcounter{footnote}{1}


\begin{titlepage}
\pagenumbering{roman}

\vspace*{-1.5cm}
\centerline{\large EUROPEAN ORGANIZATION FOR NUCLEAR RESEARCH (CERN)}
\vspace*{1.5cm}
\noindent
\begin{tabular*}{\linewidth}{lc@{\extracolsep{\fill}}r@{\extracolsep{0pt}}}
\ifthenelse{\boolean{pdflatex}}
{\vspace*{-2.7cm}\mbox{\!\!\!\includegraphics[width=.14\textwidth]{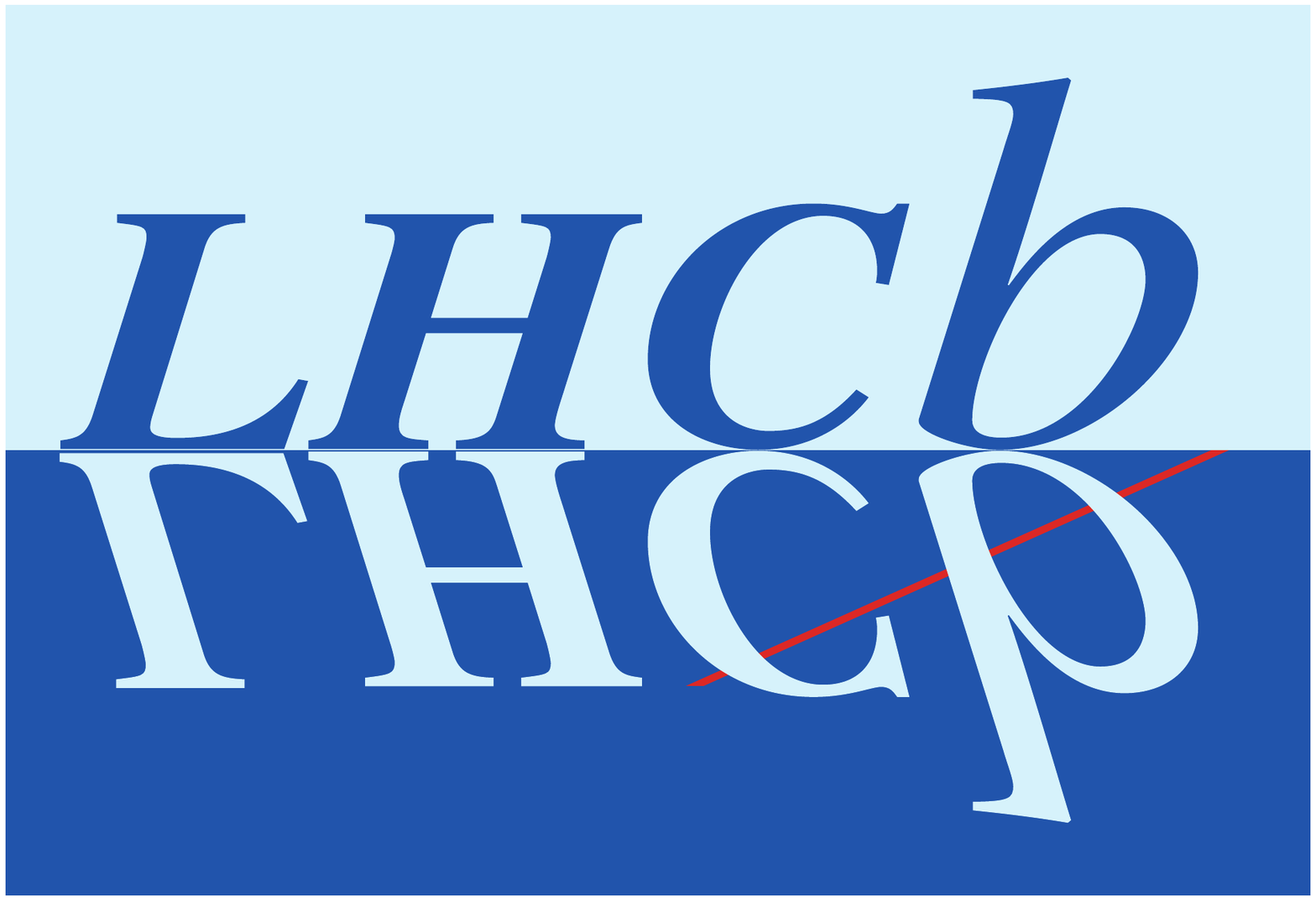}} & &}%
{\vspace*{-1.2cm}\mbox{\!\!\!\includegraphics[width=.12\textwidth]{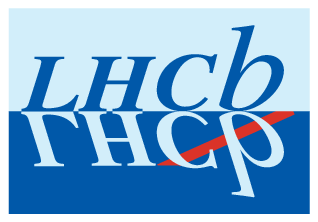}} & &}%
\\
 & & CERN-EP-2016-081 \\  
 & & LHCb-PAPER-2016-008 \\  
 & & April 5, 2016 \\ 
 & & \\
\end{tabular*}

\vspace*{1.0cm}

{\normalfont\bfseries\boldmath\huge
\begin{center}
  Measurement of the mass and lifetime of the $\Omegab$ baryon
\end{center}
}

\vspace*{0.5cm}

\begin{center}
The LHCb collaboration\footnote{Authors are listed at the end of this paper.}
\end{center}

\vspace{\fill}

\begin{abstract}
  \noindent
A proton-proton collision data sample, corresponding to an integrated luminosity of 3\invfb collected by LHCb at
$\sqrt{s}=7$ and 8~\tev, is used to reconstruct $63\pm9$ $\Omegab\to\Omegac\pim$, $\Omegac\to p\Km\Km\pip$ decays.
Using the $\Xibm\to\Xicz\pim$, $\Xicz\to p\Km\Km\pip$ decay mode for calibration,
the lifetime ratio and absolute lifetime of the $\Omegab$ baryon are measured to be
\begin{align*}
\frac{\tauOmb}{\tauXib} &= 1.11\pm0.16\pm0.03, \\ 
\tauOmb &= 1.78\pm0.26\pm0.05\pm0.06~{\rm ps},
\end{align*}
\noindent where the uncertainties are statistical, systematic and from the calibration mode (for $\tau_{\Omegab}$ only).
A measurement is also made of the mass difference, $m_{\Omegab}-m_{\Xibm}$, and the corresponding 
$\Omegab$ mass, which yields
\begin{align*}
\mOmb-\mXib &= 247.4\pm3.2\pm0.5~\mevcc, \\ 
\mOmb &= 6045.1\pm3.2\pm 0.5\pm0.6~\mevcc. 
\end{align*}
\noindent These results are consistent with previous measurements.
\end{abstract}

\vspace*{1.0cm}

\begin{center}
  Published in Phys.~Rev.~D93, 092007 (2016)
\end{center}

\vspace{\fill}

{\footnotesize 
\centerline{\copyright~CERN on behalf of the \lhcb collaboration, licence \href{http://creativecommons.org/licenses/by/4.0/}{CC-BY-4.0}.}}
\vspace*{2mm}

\end{titlepage}


\newpage
\setcounter{page}{2}
\mbox{~}
\cleardoublepage


\renewcommand{\thefootnote}{\arabic{footnote}}
\setcounter{footnote}{0}

\cleardoublepage


\pagestyle{plain} 
\setcounter{page}{1}
\pagenumbering{arabic}


%


\section{Introduction}
Measurements of the lifetimes of beauty baryons provide an important test of Heavy Quark Effective Theory 
(HQET)~\cite{Khoze:1983yp,Bigi:1991ir,Bigi:1992su,Blok:1992hw,Blok:1992he,Neubert:1997gu,Uraltsev:1998bk,Bellini:1996ra},
in which it is predicted that the decay width is dominated by the weak decay of the heavy $b$ quark. 
The large samples of $b$ baryons collected by LHCb have led to greatly improved measurements
of their lifetimes~\cite{LHCb-PAPER-2014-003,LHCb-PAPER-2014-010,LHCb-PAPER-2014-021,LHCb-PAPER-2014-048}, which
are in good agreement with HQET predictions. In particular, the lifetime of the 
$\Lb$ baryon is now measured to a precision of better than 1\%~\cite{PDG2014}, and those of the $\Xibz$ and $\Xibm$ to 
about 3\%~\cite{PDG2014,LHCb-PAPER-2014-048}.
Within HQET it is expected that the lifetimes of weakly-decaying $b$ baryons
follow the hierarchy $\tauOmb\simeq\tauXib>\tauXibz\approx\tauLb$~\cite{Bigi:1995jr,Cheng:1997xba,Ito:1997qq},
and thus far, the measured lifetimes respect this pattern within the uncertainties. 
However, the uncertainty on the measured lifetime of the $\Omegab$
baryon is too large to fully verify this prediction. The single best measurement to date of the
$\Omegab$ lifetime is $1.54^{+0.26}_{-0.21}\pm0.05$~ps~\cite{LHCb-PAPER-2014-010} by the LHCb experiment, 
based on a sample of $58\pm8$ reconstructed $\Omegab\to\jpsi\Omegam$ decays, 
with $\jpsi\to\mup\mun$, $\Omegam\to\Lz\Km$ and $\Lz\to p\pim$. 
Larger samples are needed to reduce the statistical uncertainty.

Improved knowledge of the $\Omegab$ mass would provide tighter experimental constraints for tests
of lattice quantum chromodynamics (QCD) and QCD-inspired models, which aim to accurately predict the masses of 
hadrons~\cite{amsler}.
The two most recent measurements of the $\Omegab$ mass, by the LHCb~\cite{LHCb-PAPER-2012-048} and 
CDF~\cite{Aaltonen:2014wfa} collaborations are in agreement, but an 
earlier measurement by the D0 collaboration~\cite{Abazov:2008qm} is larger by about 10 standard deviations.

In this paper, we report measurements of the mass and lifetime of the $\Omegab$ baryon using the decay mode
$\Omegab\to\Omegac\pim$, where $\Omegac\to p\Km\Km\pip$. (Charge-conjugate processes are implied throughout.) 
The only prior evidence of the $\Omegab\to\Omegac\pim$ decay has been in the $\Omegac\to\Omegam\pip$ mode,
with a signal of 4 events (3.3$\sigma$ significance)~\cite{Aaltonen:2014wfa}.
The $\mbox{\Omegac\to\proton\Km\Km\pip}$ decay mode is Cabibbo suppressed and is yet to 
be observed. However, it has the advantage of a larger acceptance in the LHCb detector compared to decay modes
with hyperons in the final state. For example, the yield of $\Xibm$ decays reconstructed
using $\mbox{\Xibm\to\Xicz\pim}$, $\mbox{\Xicz\to\proton\Km\Km\pip}$ decays~\cite{LHCb-PAPER-2014-048} is
about six times larger than that obtained using $\mbox{\Xibm\to\jpsi\Xim}$ decays~\cite{LHCb-PAPER-2014-010}, 
where $\mbox{\Xim\to\Lz\pim}$ and $\mbox{\Lz\to\proton\pim}$.

The mass and lifetime measurements are calibrated with respect to those of the $\Xibm$ baryon,
reconstructed in the $\Xibm\to\Xicz\pim$, $\Xicz\to p\Km\Km\pip$ decay mode. The mass and lifetime of the
$\Xibm$ are measured to be $\mXib=5797.72\pm0.55\mevcc$ and $\tauXib=1.599\pm0.041\pm0.022$~ps~\cite{LHCb-PAPER-2014-048}, 
respectively; the measurements are of sufficiently high precision that they do not represent a limiting uncertainty in the
$\Omegab$ measurements presented here. The two quantities that
are measured are the mass difference, $\delta m = \mOmb-\mXib$, and the lifetime ratio
$\tauOmb/\tauXib$. The identical final states and similar energy release in the 
$b$ and $c$ baryon decays lead to a high degree of cancellation of the systematic uncertainties on these quantities.
Throughout this article, we use $X_b$ ($X_c$) to refer to either a $\Xibm$ ($\Xicz$) or $\Omegab$ ($\Omegac$) baryon.

\section{Detector and simulation}
The measurements use proton-proton ($pp$) collision data samples, collected by the LHCb experiment,
corresponding to an integrated luminosity of 3.0\invfb, of which 1.0\invfb was recorded at 
a center-of-mass energy of 7\tev and 2.0\invfb at 8\tev.  
The \lhcb detector~\cite{Alves:2008zz,LHCb-DP-2014-002} is a single-arm forward
spectrometer covering the \mbox{pseudorapidity} range $2<\eta <5$,
designed for the study of particles containing \bquark or \cquark
quarks. The detector includes a high-precision tracking system
consisting of a silicon-strip vertex detector surrounding the $pp$
interaction region, a large-area silicon-strip detector located
upstream of a dipole magnet with a bending power of about
$4{\mathrm{\,Tm}}$, and three stations of silicon-strip detectors and straw
drift tubes placed downstream of the magnet.
The tracking system provides a measurement of momentum of charged particles with
a relative uncertainty that varies from 0.5\% at low momentum to 1.0\% at 200\gevc.
The minimum distance of a track to a primary vertex (PV), the impact parameter (IP), is measured with a resolution of $(15+29/\pt)\mum$,
where \pt is the component of the momentum transverse to the beam, in\,\gevc.
Different types of charged hadrons are distinguished using information
from two ring-imaging Cherenkov detectors.
Photons, electrons and hadrons are identified by a calorimeter system consisting of
scintillating-pad and preshower detectors, an electromagnetic
calorimeter and a hadronic calorimeter. Muons are identified by a
system composed of alternating layers of iron and multiwire
proportional chambers.

The online event selection is performed by a trigger~\cite{LHCb-DP-2012-004},
which consists of a hardware stage, based on information from the calorimeter and muon
systems, followed by a software stage, which applies a full event
reconstruction. The software trigger requires a two-, three- or four-track
secondary vertex with a large \pt sum of the tracks
and a significant displacement from the primary $pp$
interaction vertices. At least one particle should have $\pt>1.7\gevc$ and 
be inconsistent with coming from any of the PVs. The signal candidates 
are required to pass a multivariate software trigger
selection algorithm~\cite{BBDT}. 

Proton-proton collisions are simulated using
\pythia~\cite{Sjostrand:2006za,*Sjostrand:2007gs} with a specific \lhcb
configuration~\cite{LHCb-PROC-2010-056}.  Decays of hadronic particles
are described by \evtgen~\cite{Lange:2001uf}, in which final-state
radiation is generated using \photos~\cite{Golonka:2005pn}. The
interaction of the generated particles with the detector, and its
response, are implemented using the \geant toolkit~\cite{Allison:2006ve, *Agostinelli:2002hh} as described in
Ref.~\cite{LHCb-PROC-2011-006}. The $\Xicz\to p\Km\Km\pip$ and $\Omegac\to p\Km\Km\pip$ decays are
modeled as an equal mixture of $X_c\to p\Km\Kstarzb,~\Kstarzb\to\Km\pip$ and $X_c\to p\Km\Km\pip$ (nonresonant) decays; 
this composition reproduces well the only clear structure in these decays, a $\Kstarzb$ peak in the $\Km\pip$ mass distribution.

\section{Candidate selection}

Candidate $X_c\to p\Km\Km\pip$ decays are formed by combining four tracks consistent with this decay chain,
and requiring a good quality vertex fit. In forming the $X_c$ candidate, 
each particle must be significantly detached from all PVs in the event, have $\pt$ greater than 100\mevc, and have
particle identification (PID) information consistent with the decay hypothesis.
The PID requirements on the proton and the kaon candidates have a combined efficiency of 70\% on signal, while reducing the
combinatorial background by a factor of 3.5. 

Candidate $X_b$ baryons are formed by combining an $X_c$ candidate with a $\pim$ candidate.
For each $X_b$ and PV pair in an event, a quantity $\chisqip(X_b)$ is computed, defined as the increase in $\chi^2$ 
when the $X_b$ candidate is included as an additional particle in the PV fit. The $X_b$ candidate is assigned to the
PV with the smallest value of $\chisqip(X_b)$, and it is required to be significantly displaced from that PV. 
The invariant mass $M(p\Km\Km\pip)$ is required to lie in the range
2461--2481\mevcc and 2685--2705\mevcc for $\Xicz$ and $\Omegac$ signal candidates, respectively; these intervals cover a 
mass region that represents about $\pm$2.5 and $\pm$2.0 times the expected mass resolution. The tighter requirement on 
the $\Omegac$ candidates is used because of a lower signal-to-background ratio. Candidates for which the $p\Km\Km\pip$ mass is
outside the signal region are also used to model the $X_c$ combinatorial background contribution to the signal sample.
To suppress combinatorial background, candidate $X_b$ decays are required to have a reconstructed decay time
larger than 0.2~ps, which is about five times the decay time resolution for these decays.

To further improve the signal-to-background ratio, a multivariate analysis is employed,
based on a boosted decision tree (BDT) algorithm~\cite{Breiman, AdaBoost}
implemented within the TMVA package~\cite{Hocker:2007ht}. Simulated $\Xibm$ and $\Omegab$ decays are used to represent 
the signal distributions, and background events are taken from the signal sidebands in data. The sidebands consist of
events that are close in mass to the $X_b$ signal region, but have either the $p\Km\Km\pip$ or $X_c\pim$ mass inconsistent with the 
known $X_c$ or $X_b$ masses. Independent training and test samples are used to ensure that the BDT is not overtrained.

A total of 18 discriminating variables are used to help differentiate signal and background candidates, including:
the $X_b$ decay vertex fit $\chisq$; the $\chisqip$ of the $X_b$, $X_c$ and final-state decay products; 
the consistency of the candidate with being produced at one of the PVs in the event; the \pt of the decay products;
and the PID information on the proton and two kaons. Due to differences in the PID information between simulation and data, 
the distributions of PID variables for signal are taken from 
$\mbox{\Dstarp\to\Dz\pip}$ with $\Dz\to\Km\pip$, $\Lz\to p\pim$ and $\Lc\to p\Km\pip$ decays in 
data~\cite{LHCb-DP-2012-003}, and are reweighted to account for differences in kinematics between the control and
signal samples. The output of the training is a single discriminating 
variable that ranges from $-1$ to $1$. For convenience, the output value is also referred to as BDT. 

The BDT requirement is chosen to maximize the figure of merit $N_S/\sqrt{N_S+N_B}$ for the $\Omegab$ signal.
Here, $N_S$ and $N_B$ are the expected signal and background yields as a function of the BDT requirement.
The chosen requirement of BDT$>$0.3 provides an expected signal (background) efficiency of about 90\% (10\%).

\section{Mass spectra and fits}

\begin{figure}[tb]
\centering
\includegraphics[width=0.98\textwidth]{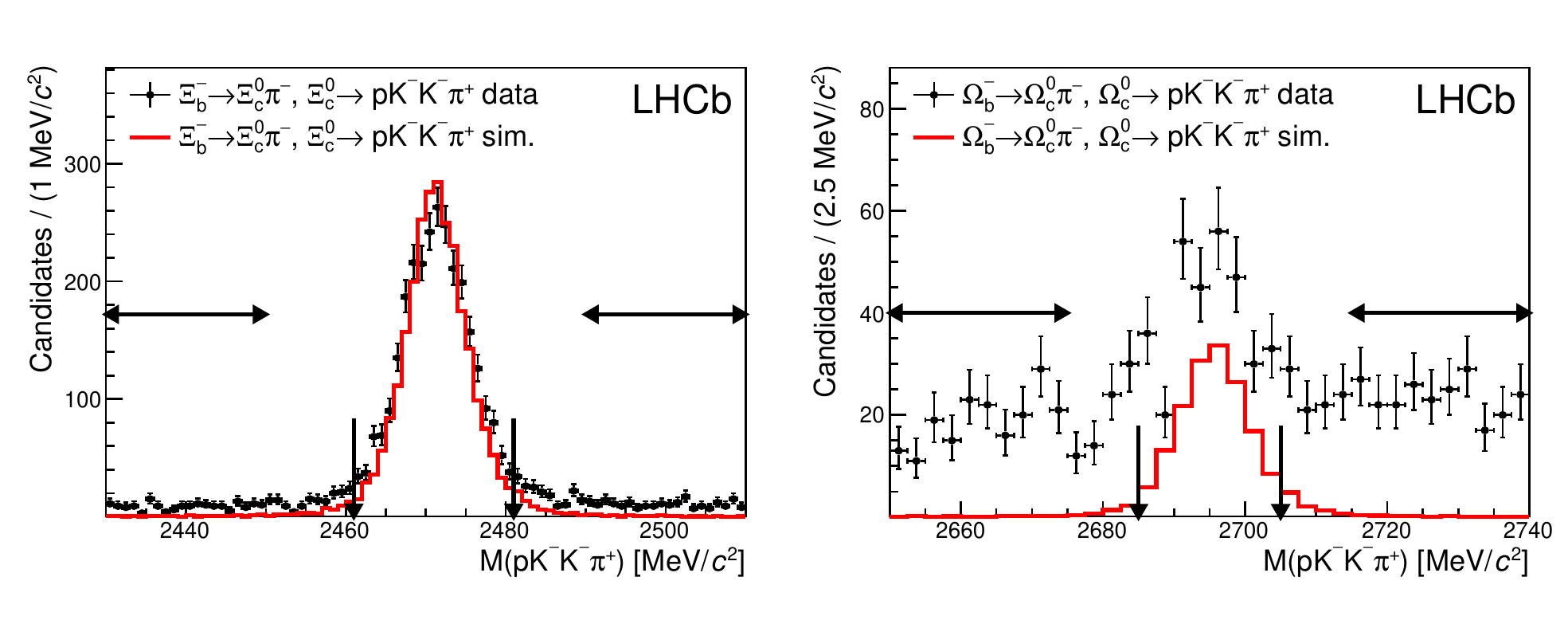}
\caption{\small{Invariant mass distribution for (left) $\Xicz\to p\Km\Km\pip$ and (right) $\Omegac\to p\Km\Km\pip$ candidates
over the full $X_b$ fit regions. The corresponding simulations (sim.) are overlaid. The vertical arrows indicate the
signal regions, and the horizontal ones show the sideband regions.}}
\label{fig:XcMassComp}
\end{figure}

The $X_c$ invariant mass spectra for $X_b$ signal candidates are shown in Fig.~\ref{fig:XcMassComp}. All candidates
within the regions contributing to the $\Omegab$ mass fit, 5420--6380\mevcc, and the $\Xibm$ mass fit, 5630--6590\mevcc, are included.
The simulated distributions, normalized to the fitted number of $X_c$ signal decays in data, are overlaid. The vertical 
and horizontal arrows indicate the signal and sideband regions. 

While the overall background yields in these spectra are comparable, the signal-to-background ratio is much lower within the
$\Omegac$ candidate sample due to the lower production rate of $\Omegab$ relative to $\Xibm$ baryons, and likely a smaller $X_c\to p\Km\Km\pip$
branching fraction. Due to the very different $X_c$ background levels for the signal and calibration mode, we use the
$X_c$ sidebands to model the $X_c$ combinatorial background in the $X_b$ invariant mass spectra.

     To measure the $\Omegab$ mass and yield, the data are fitted using a simultaneous extended unbinned maximum 
likelihood fit to four $X_b$ invariant mass distributions; one pair is formed from the $X_c$ signal
regions, and the second pair comprises events taken from the $X_c$ sidebands, as indicated in Fig.~\ref{fig:XcMassComp}.

The signal shapes, determined from $\Omegab\to\Omegac\pim$ and $\Xibm\to\Xicz\pim$ simulated events, 
are each modeled by the sum of two Crystal Ball (CB) functions~\cite{Skwarnicki:1986xj} which
have a common mean value. The general forms of the two signal shapes are
\begin{align}
\mathcal{F}_{\rm sig}^{\Xibm} &= f_{\rm low}{\rm CB}_{-}(m_0,\fsigma\rsigma\sigma,\alpha_{-},N_{-}) + (1-f_{\rm low}){\rm CB}_{+}(m_0,\fsigma\sigma,\alpha_{+},N_+), \\
\mathcal{F}_{\rm sig}^{\Omegab} &= f_{\rm low}{\rm CB}_{-}(m_0+\delta m,\rsigma\sigma,\alpha_{-},N_{-}) + (1-f_{\rm low}){\rm CB}_{+}(m_0+\delta m,\sigma,\alpha_{+},N_+).
\end{align}
\noindent Several of the parameters are common in the two signal shapes, and are determined from a simultaneous fit to
the mass spectra from simulated samples of $\Omegab$ and $\Xibm$ decays.
The ${\rm CB}_{\pm}$ function represents the signal contribution with a tail toward low ($-$) or high ($+$) invariant mass.
The parameters $m_0$ and $m_0+\delta m$ represent the fitted peak mass values of the $\Xibm$ and $\Omegab$ baryons, respectively;
$\rsigma$ relates the lower CB width to the upper one; and $\fsigma$ allows for a
small difference in the mass resolution for the signal and calibration modes.
The exponential tail parameters $\alpha_{\pm}$ are common to the signal and calibration modes. 
We fix the power-law tail parameters $N_{-}=N_{+}=10$, and the fraction $f_{\rm low}=0.5$, as the simulated signal shapes are well described 
without these parameters freely varied. In fits to the data, $m_0$, $\delta m$ and $\sigma$ are left free to vary, and all other shape parameters are
fixed to the values from the simulation.

Several sources of background contribute to the invariant mass spectrum for both the signal and the calibration modes.
These include: (i) partially-reconstructed $X_b\to X_c\rho^-$ decays; (ii) misidentified $X_b\to X_c\Km$ decays;
(iii) partially-reconstructed $\Omegab\to\OmegacStar\pim$ decays ($\Omegab$ only);
(iv) random $X_c\to p\Km\Km\pip$ combinations; and (v) $X_b\to X_c\pim$ combinatorial background.
The $X_b\to X_c\rho^-$ background shape is based on simulated decays, and is parameterized by an ARGUS distribution~\cite{Albrecht:1994tb} 
convolved with a Gaussian resolution function of 16.4\mevcc fixed width, the value obtained 
from fully reconstructed $\Omegab\to\Omegac\pim$ decays in data. 
The ARGUS shape parameters are left free to vary in the fit,
as is the yield, expressed as a fraction of the $X_b\to X_c\pim$ yield. The $X_b\to X_c\Km$ background
shape is fixed based on simulation. The yield fraction $N(X_b\to X_c\Km)/N(X_b\to X_c\pim)$ is fixed to 3.1\%, which is
the product of an assumed ratio of branching fractions $\br(X_b\to X_c\Km)/\br(X_b\to X_c\pim)=7\%$, based on the
value from $\Lb$ decays~\cite{LHCb-PAPER-2013-056}, and the efficiency of the PID requirements on the $\Km$ and $\pim$.
The shape parameters used to describe these two backgrounds are common to the signal and calibration modes,
apart from an overall mass offset, which is fixed to be equal to $\delta m$. The invariant mass distribution of the
$\Omegab\to\OmegacStar\pim$ background is taken from a parametrization of the mass distribution 
obtained from a phase-space simulation~\cite{Brun:1997pa}, combined with a Gaussian smearing
based on the measured mass resolution. The yield fraction $N(\Omegab\to\Omegac\pim)/N(\Omegab\to\OmegacStar\pim)$
is freely varied in the fit to data.

The $X_c\to p\Km\Km\pip$ combinatorial background contribution is constrained by including the $X_c$ sidebands in the simultaneous fit, 
as discussed above. The shape of this background is modeled by the sum of a broad Gaussian function and an exponential shape. 
In the $X_c$ sidebands there is no indication of any
$\Xibm$ or $\Omegab$ contributions, which might result from nonresonant $X_b\to p\Km\Km\pip\pim$ decays. The shape parameters and 
yields of this background component are freely varied in the fit, but their values are common for the $X_c$ signal and sideband data samples.
A different set of parameters is used for the $\Omegab$ and $\Xibm$ decay modes. Random $X_c\pim$ combinatorial background
is described by a single exponential function with variable slope and yield.  

\begin{figure}[tb]
\centering
\includegraphics[width=0.49\textwidth]{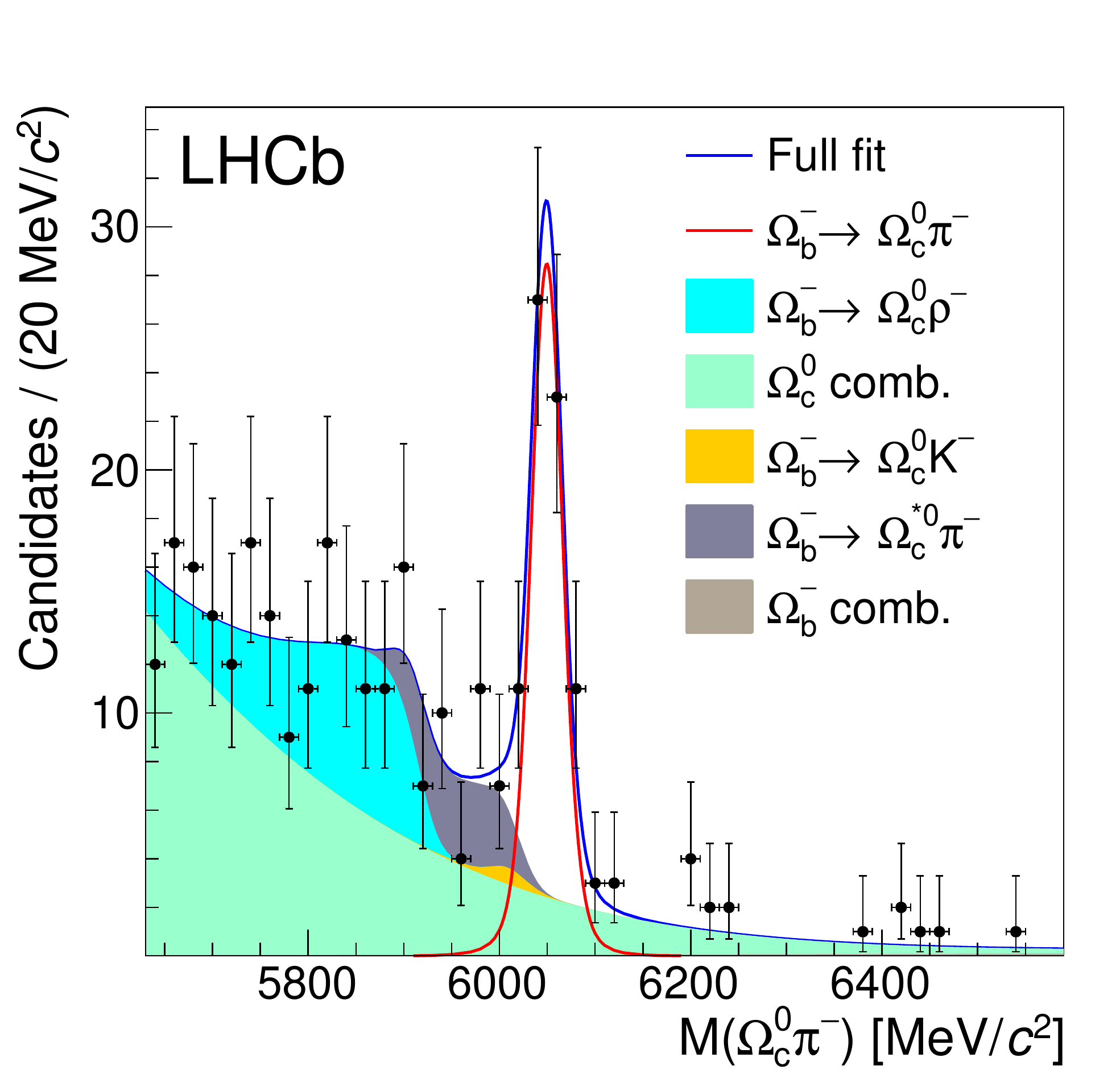}
\includegraphics[width=0.49\textwidth]{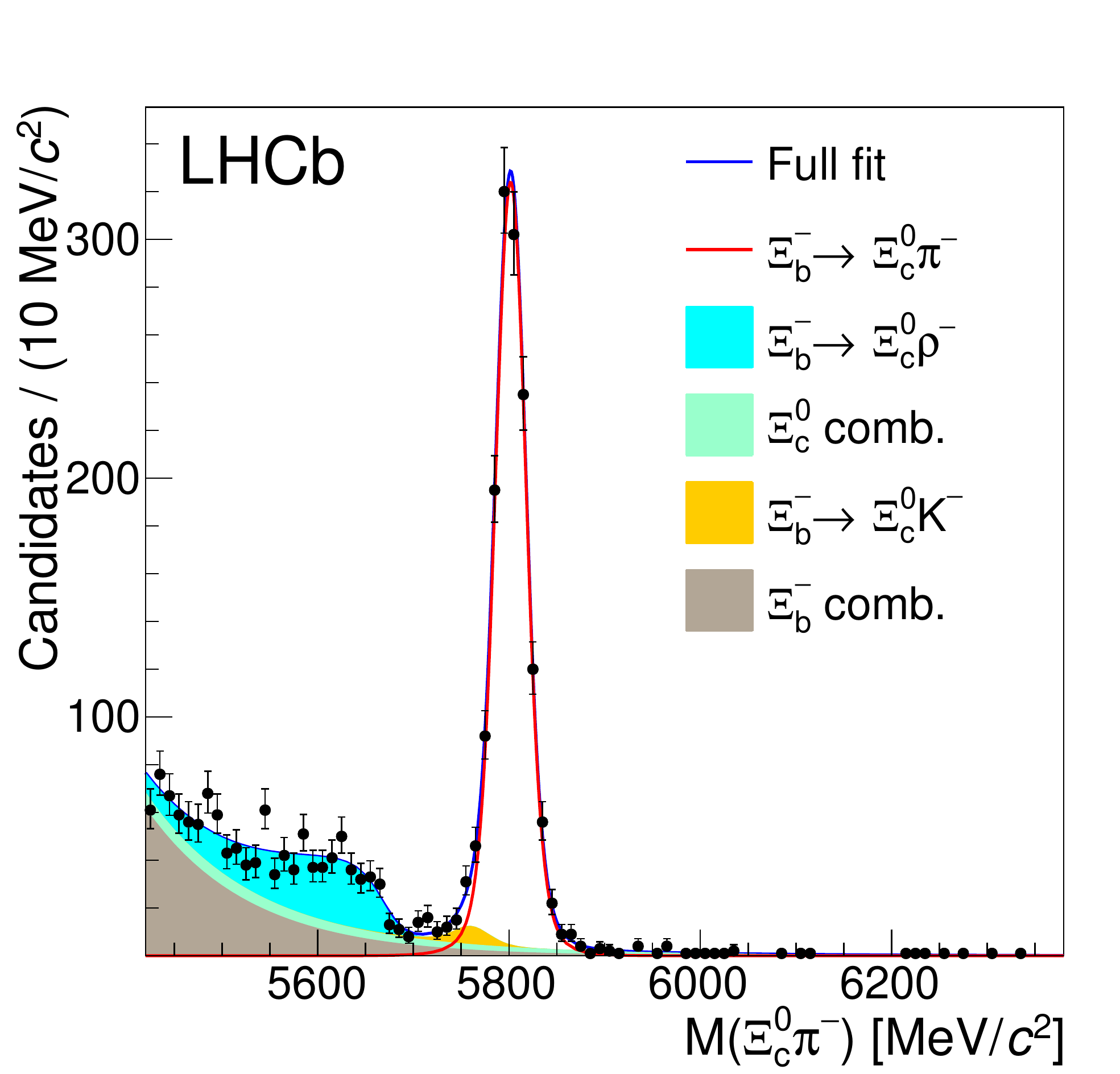}
\caption{\small{Results of the simultaneous mass fit to the signal and calibration modes. The fitted $\Omegab$ combinatorial (comb.) 
background yield is very small, and not clearly visible.}}
\label{fig:MassFitsData}
\end{figure}

The $X_b$ invariant mass spectra with the fits overlaid are shown in Fig.~\ref{fig:MassFitsData} for the $X_c$ signal 
regions. The fitted yields are
$62.6\pm9.0$ and $1384\pm39$ for the $\Omegab\to\Omegac\pim$ and $\Xibm\to\Xicz\pim$ modes, respectively.
The $\Omegab\to\Omegac\pim$, $\Omegac\to p\Km\Km\pip$ decay is observed for the first time with large 
significance, about 10 standard deviations based on Wilks's theorem~\cite{Wilks:1938dza}.
The yield of $\Omegab\to\Omegac\pim$ decays is comparable to that obtained in $\Omegab\to\jpsi\Omegam$ 
decays~\cite{LHCb-PAPER-2014-010}. The mass difference is measured to be $\delta m=247.7\pm3.0$\mevcc, where the uncertainty 
is statistical only.

\section{$\Omegab$ lifetime}
To measure the $\Omegab$ lifetime, the data from the signal and calibration modes are divided into four
bins of $X_b$ decay time:  0.0--1.5~ps, 1.5--2.5~ps, 2.5--4.0~ps, and 4.0--12.0~ps. The decay time binning was chosen
based on pseudoexperiments which replicate the yields of events in data as a function of decay time for the signal and calibration 
modes. Several binning schemes were investigated and the one above minimizes the systematic uncertainty on the lifetime due
to the small $\Omegab$ sample size.

\begin{figure}[tb]
\centering
\includegraphics[width=0.45\textwidth]{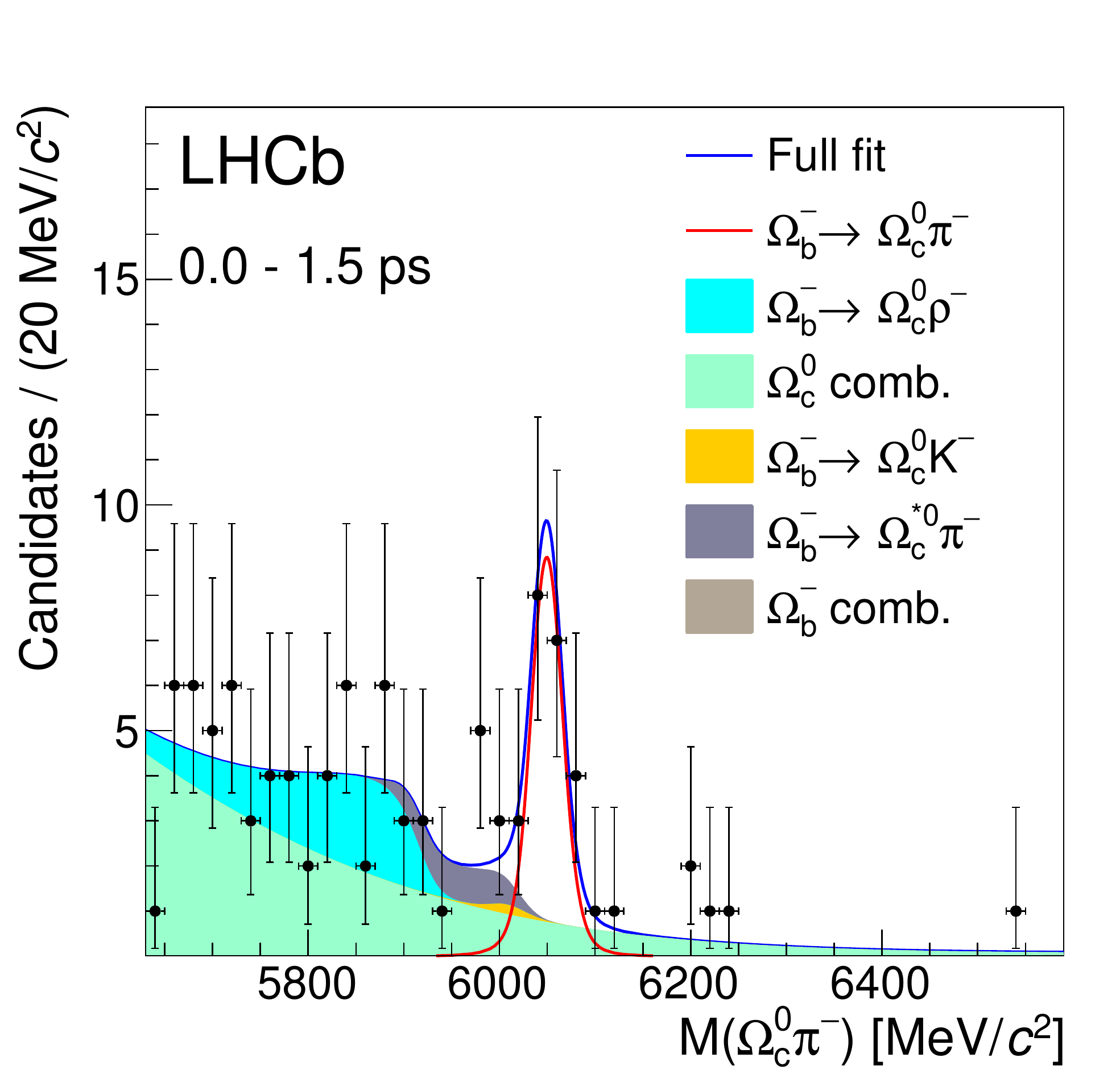}
\includegraphics[width=0.45\textwidth]{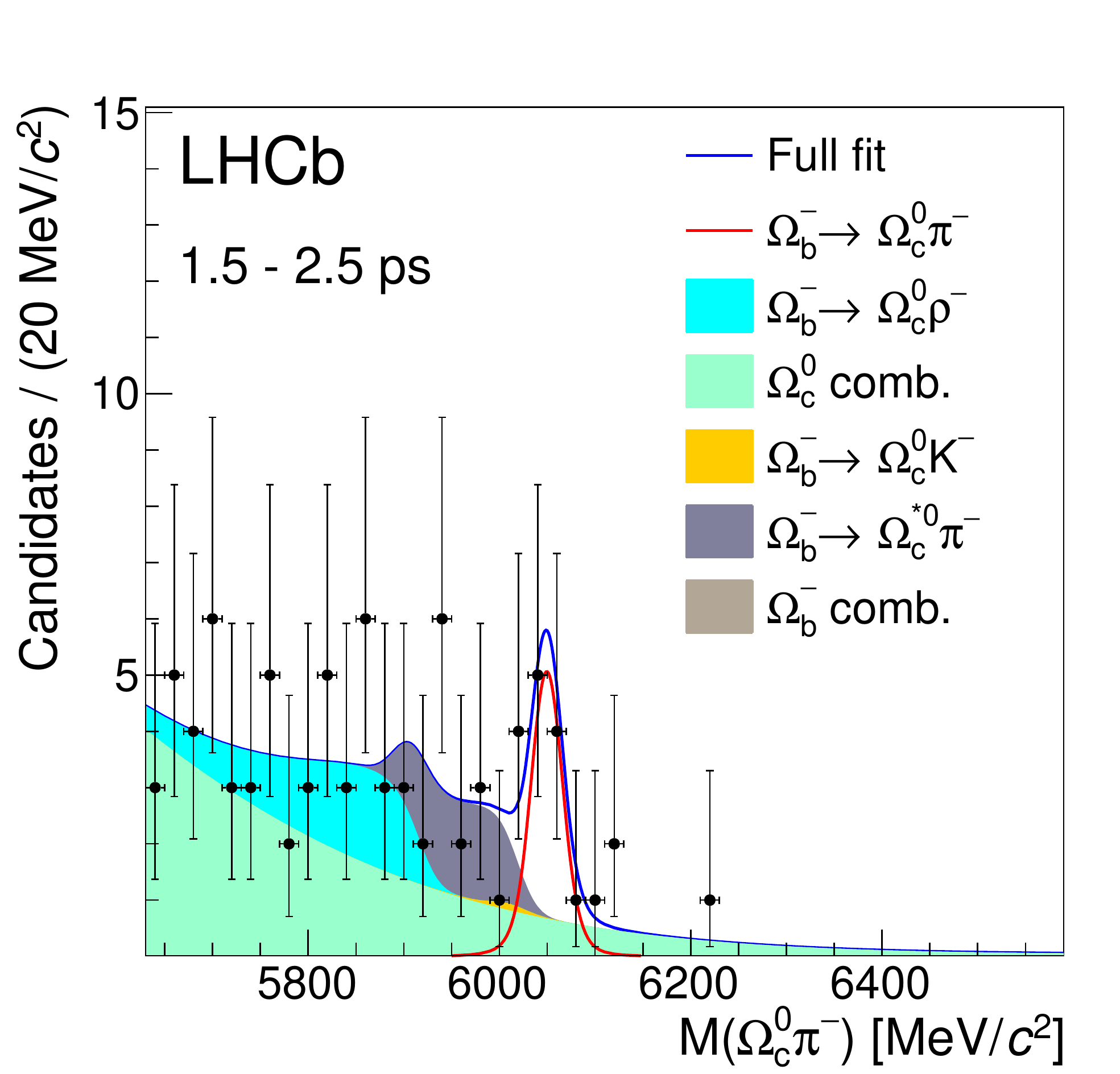}
\includegraphics[width=0.45\textwidth]{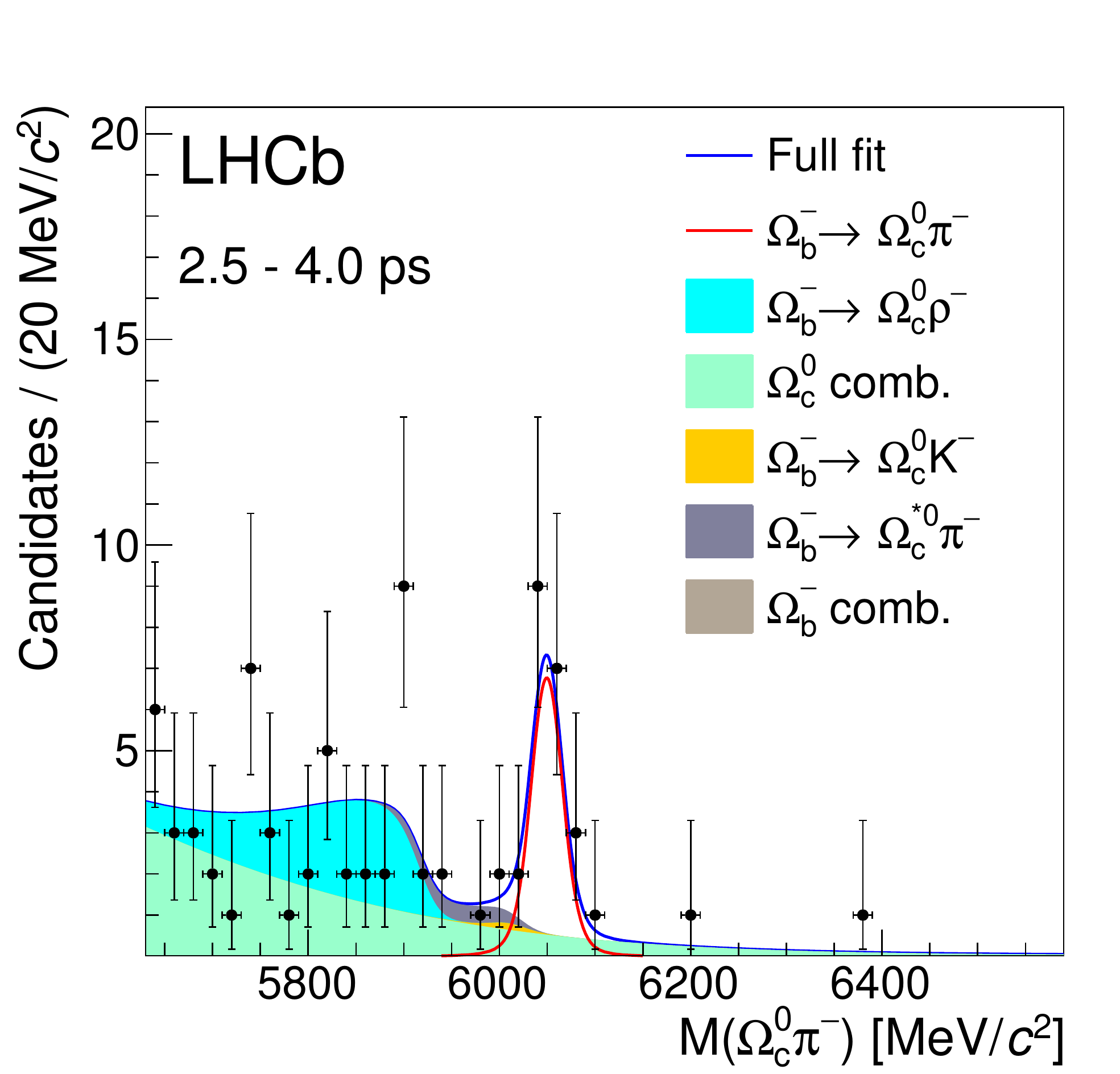}
\includegraphics[width=0.45\textwidth]{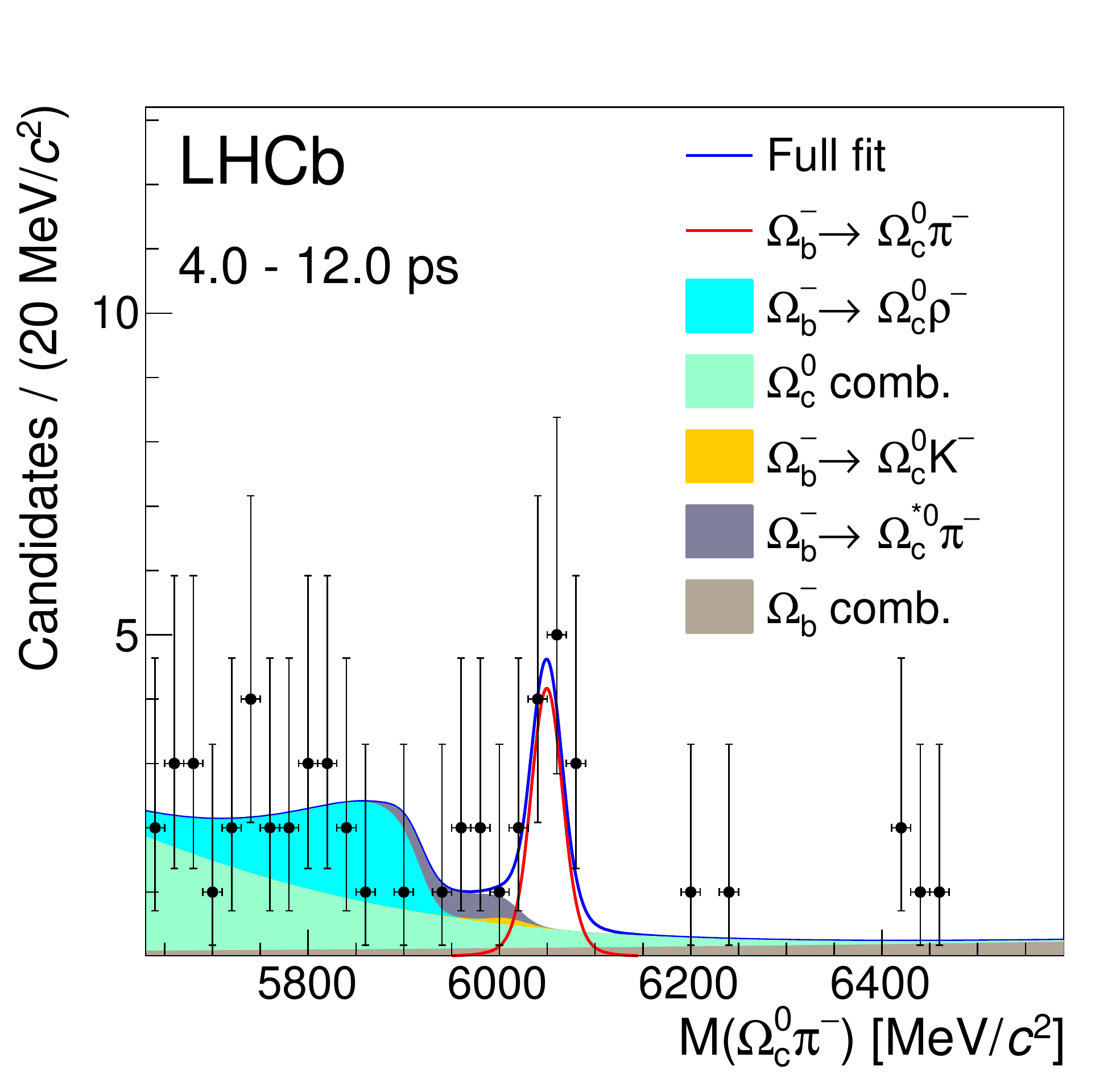}
\caption{\small{Results of the simultaneous mass fit to the $\Omegab$ signal in the four decay time bins, as
indicated in each plot.}}
\label{fig:OmbTimeBins}
\end{figure}

\begin{figure}[tb]
\centering
\includegraphics[width=0.45\textwidth]{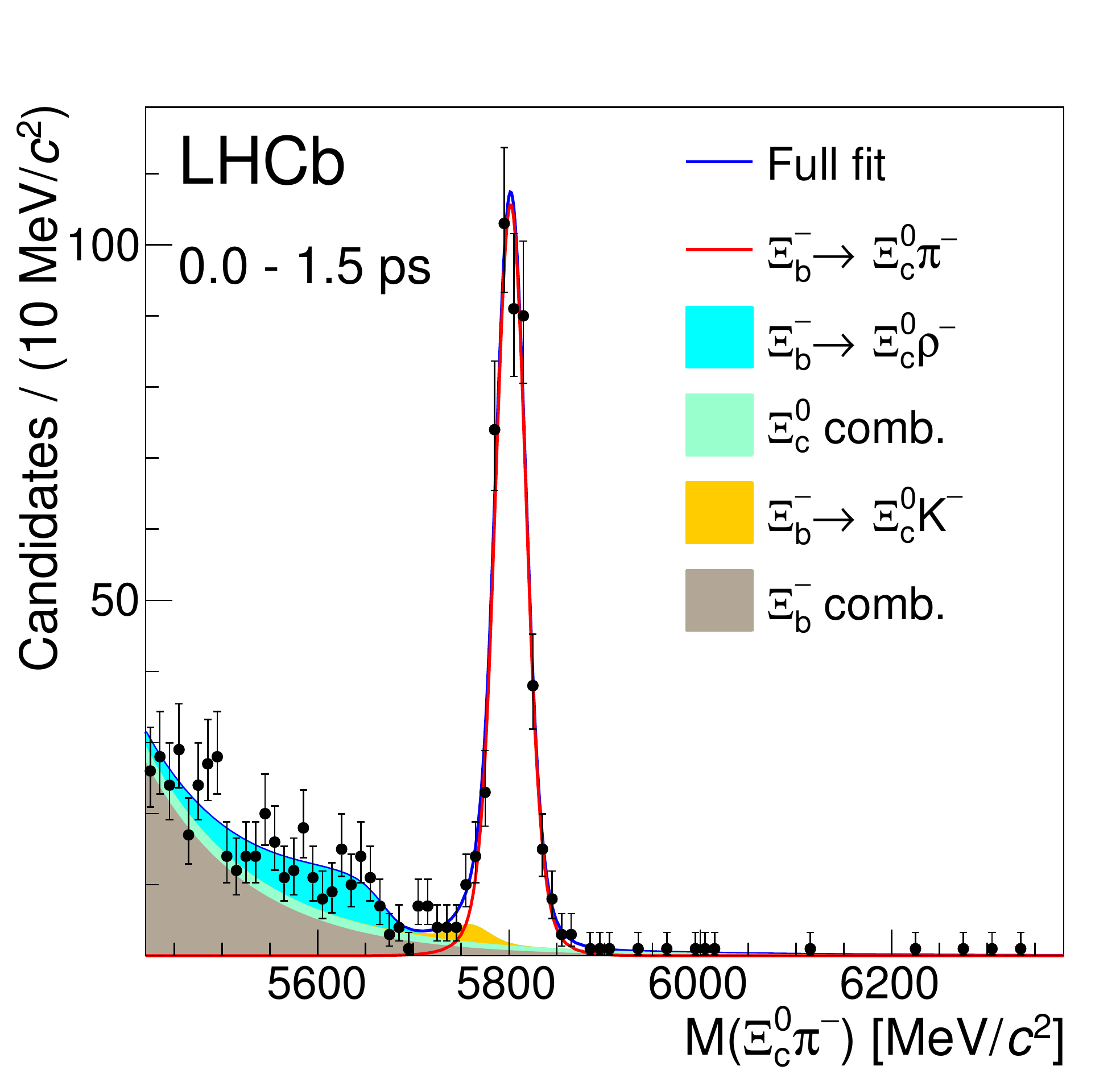}
\includegraphics[width=0.45\textwidth]{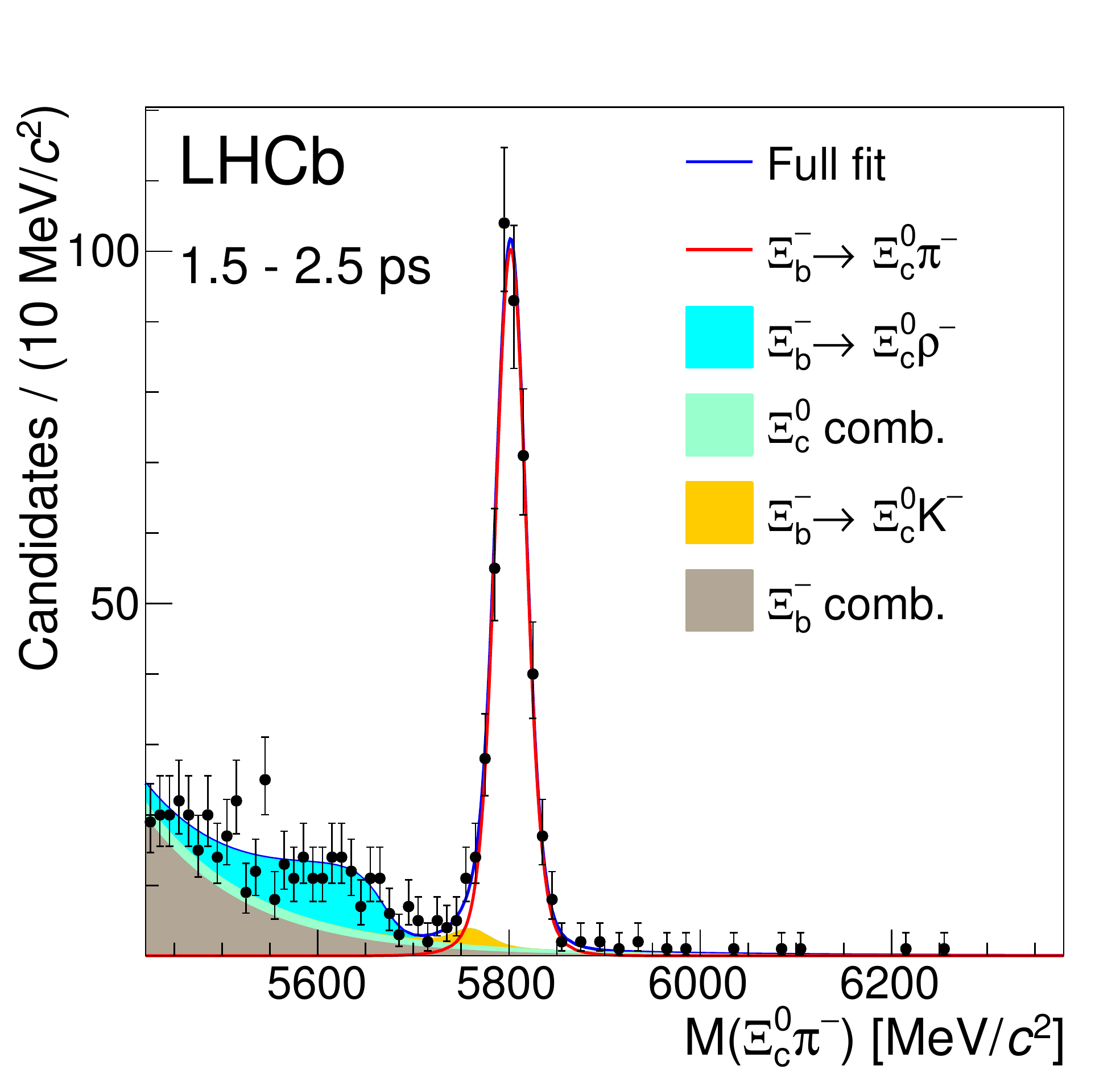}
\includegraphics[width=0.45\textwidth]{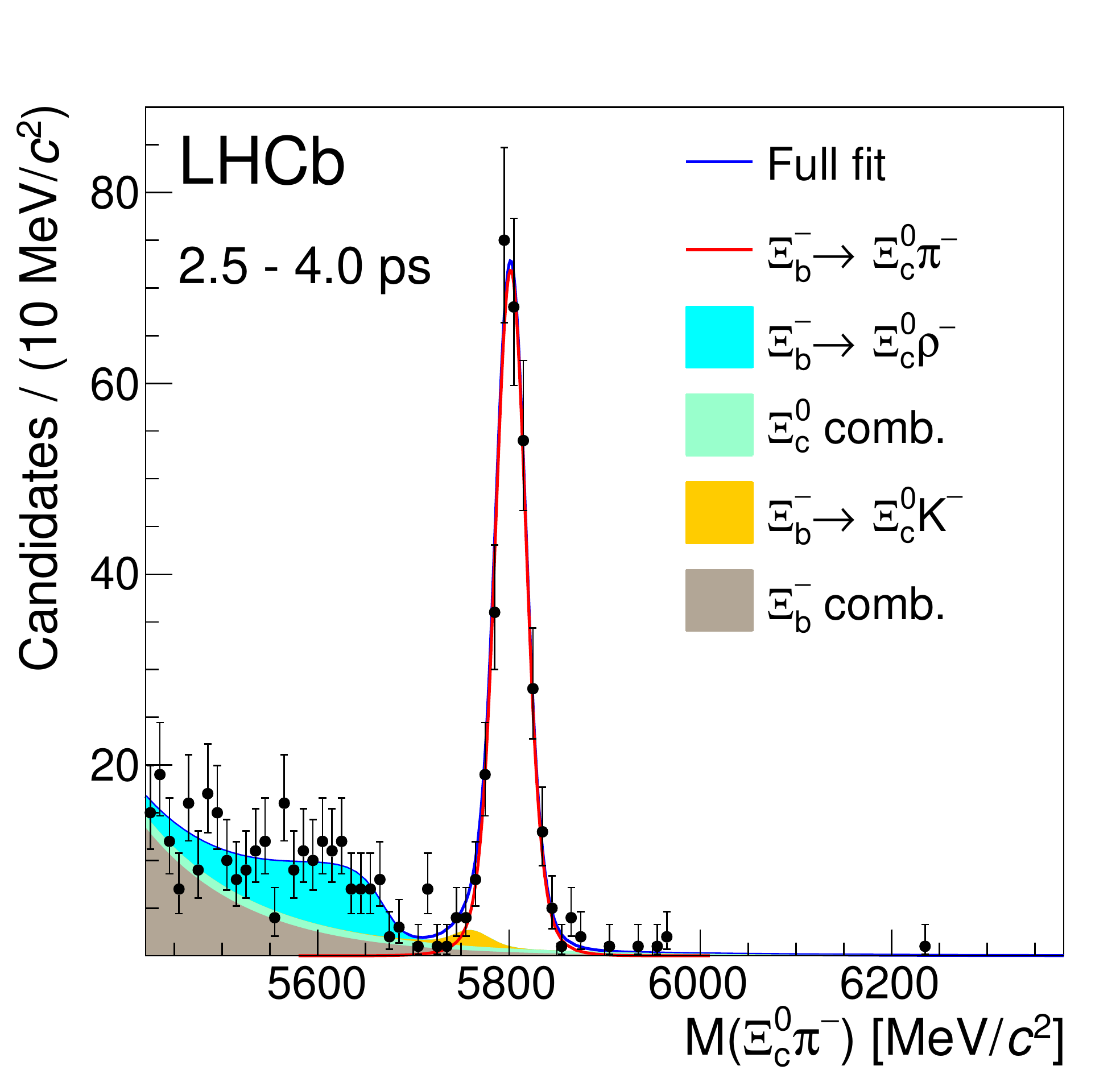}
\includegraphics[width=0.45\textwidth]{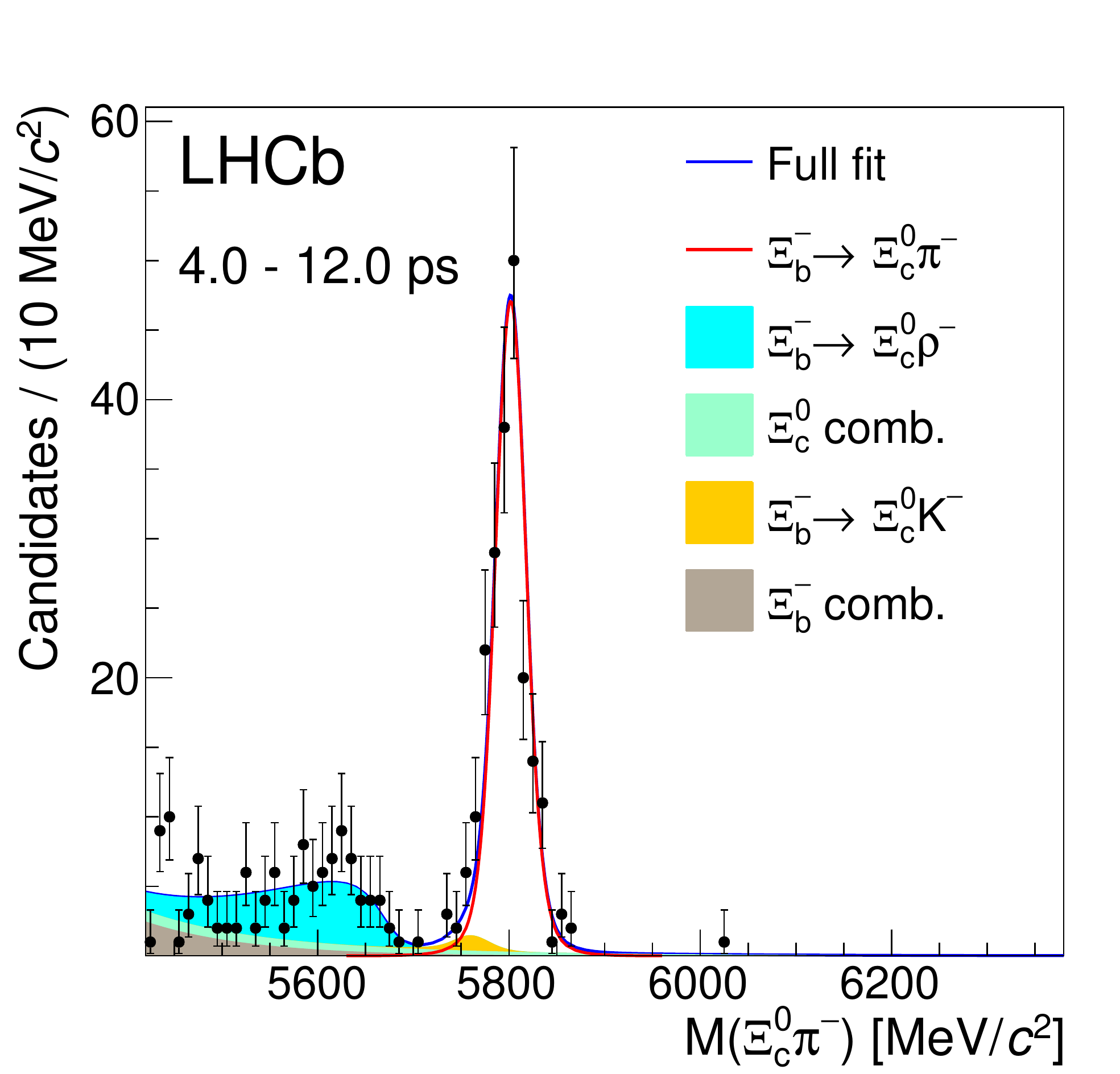}
\caption{\small{Results of the simultaneous mass fit to the $\Xibm$ signal in the four decay time bins, as
indicated in each plot.}}
\label{fig:XibTimeBins}
\end{figure}

The yields in each decay time bin in data are determined by repeating the mass fit for each decay time bin, allowing the signal and 
background yields to vary freely. All shape parameters are fixed to the values 
obtained from the fit to the whole data sample, since simulations show that they do not depend on the decay time.
The results of the fits to the individual decay time bins are shown in Figs.~\ref{fig:OmbTimeBins} and~\ref{fig:XibTimeBins}
for the signal and calibration modes. The yields are presented in Table~\ref{tab:YieldInTimeBins}.
\begin{table*}[tb]
\begin{center}
\caption{\small{Results of the fit to data for each decay time bin, and the relative efficiency. The uncertainties are statistical only.}}
\begin{tabular}{lccc}
Decay time bin (ps) & $\Omegab$ yield & $\Xibm$ yield & $\eff(\Xibm)/\eff(\Omegab)$ \\
\hline
0.0--1.5 & $20.8\pm4.8$ & $450\pm21$ & $1.10\pm0.03$ \\
1.5--2.5 & $12.0\pm3.7$  & $427\pm21$ & $1.11\pm0.04$ \\
2.5--4.0 & $17.7\pm 4.2$ & $305\pm17$ & $1.02\pm0.04$ \\
4.0--12.0 & $10.5\pm3.3$  & $201\pm14$ & $1.03\pm0.05$ \\
\end{tabular}
\label{tab:YieldInTimeBins}
\end{center}
\end{table*}

The relative efficiency in each bin is determined using simulated events. The efficiency-corrected yield ratio is then
\begin{align}
\frac{N_{\Omegab\to\Omegac\pim}(t)}{N_{\Xibm\to\Xiz\pim}(t)} = A\exp{(\kappa t)},
\end{align}
\noindent where $A$ is a calibration factor, and 
\begin{align}
\kappa\equiv 1/\tauXib - 1/\tauOmb.
\end{align}
The value of $\kappa$ is obtained by fitting an exponential function to the efficiency-corrected ratio of yields, 
which in turn allows $\tauOmb$ to be determined.
The efficiencies for the signal and normalization modes are
expressed as the fraction of generated signal decays with true decay time in bin $i$, which have a reconstructed decay
time also in bin $i$. When defined in this way, effects of time resolution and selection requirements are accounted for, 
and the corrected signal and calibration mode yields are exponential in nature.
The relative efficiencies after all selection requirements are given in Table~\ref{tab:YieldInTimeBins}.

\begin{figure}[t]
\centering
\includegraphics[width=0.98\textwidth]{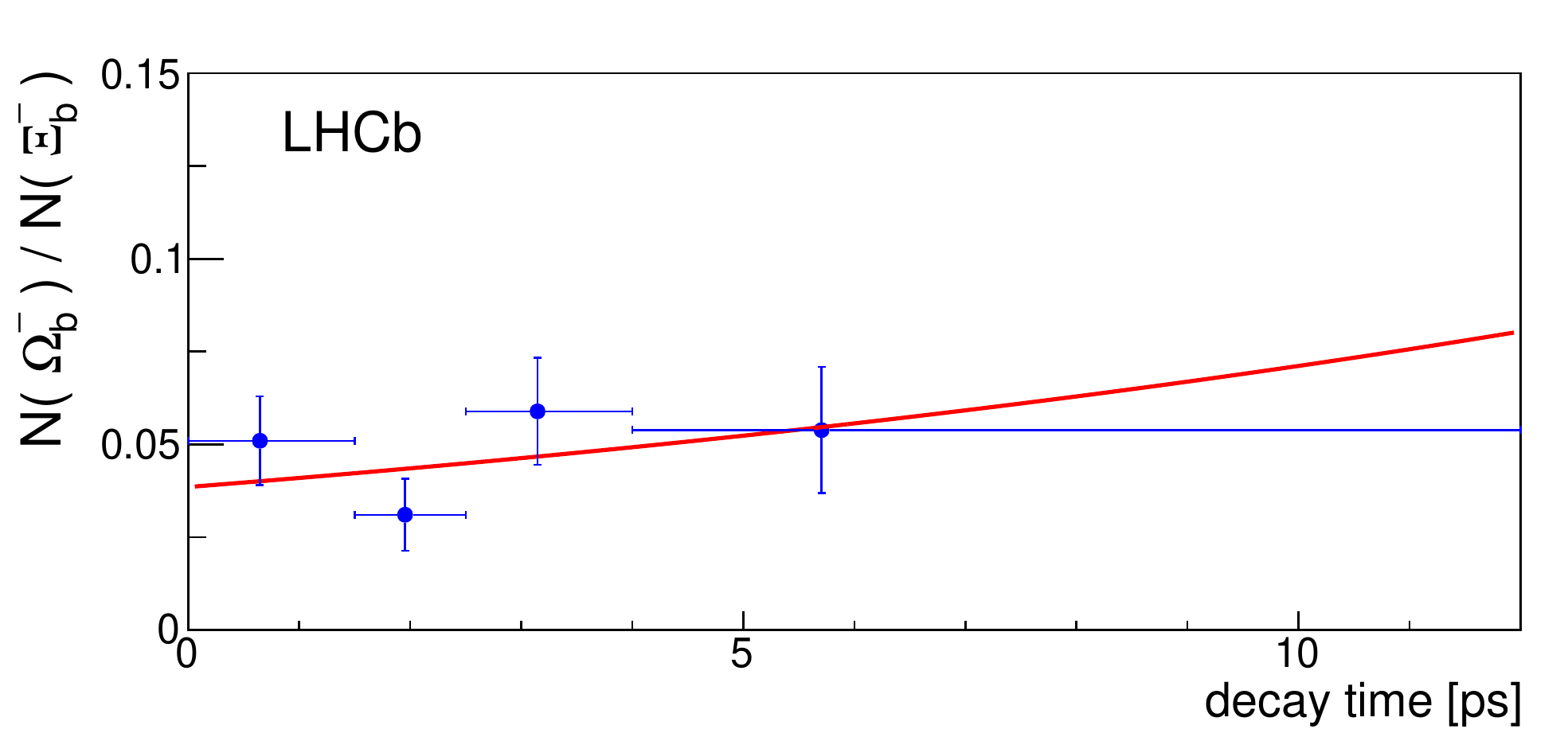}
\caption{\small{Corrected signal yield ratio as a function of decay time, along with a fit to an exponential function. The
horizontal bars indicate the bin sizes, and are not an indication of the uncertainty.}}
\label{fig:CorrYieldRatio}
\end{figure}

The efficiency ratio is consistent with having no dependence on the decay time, as expected from the
similarity of the two decay modes. 
\indent The efficiency-corrected yield ratio as a function of decay time is shown in Fig.~\ref{fig:CorrYieldRatio}, along with a $\chi^2$ fit to the
data using an exponential function. The 
position of the points along the decay time axis is determined by taking the
average value within the bin, assuming an exponential decay time distribution with $\tau=1.60$~ps. 
From the fitted value of $\kappa=0.053\pm0.085$~ps$^{-1}$ and the measured value of the $\Xibm$ lifetime,
the lifetime ratio is found to be
\begin{align}
\frac{\tauOmb}{\tauXib} &= \frac{1}{1-\kappa\tauXib} =  1.09\pm0.16,
\label{eq:ltratio}
\end{align}
\noindent where the uncertainty is statistical only.

\section{Systematic uncertainties}
A number of systematic uncertainties are evaluated, and are summarized in Table~\ref{tab:SummaryOfShifts}. 
Most of the systematic uncertainties are estimated by modifying each
fixed input or function, and taking the difference with respect to the nominal value as the systematic uncertainty.
The signal shape uncertainty is determined by changing the description to the sum of two Gaussian functions and repeating 
the analysis. The nominal $X_c$ combinatorial background shape is changed from the sum of a Gaussian shape and an exponential function to a single exponential
distribution. The sensitivity to the $\Omegab\to\OmegacStar\pim$ shape description is investigated by varying the shape parameters
obtained from the simulation to account for the uncertainty on the mass resolution, as well as 
using a different function to parametrize the simulation. The uncertainty on the yield of misidentified 
$X_b\to X_c\Km$ decays is quantified by varying the fractional contribution by $\pm$30\% relative to the
nominal value, to allow for uncertainty in the $X_b\to X_c\Km$ branching fractions amongst these modes and
for uncertainty in the PID efficiencies. The relative
efficiency is obtained from simulation. However, the BDT performance in data is slightly worse than in simulation, so to estimate 
a potential bias in the lifetime ratio, we re-evaluate the relative efficiency with a BDT$>$0.6 requirement, while keeping the nominal requirement
on the data. This larger value was chosen since it provides equal efficiency of the BDT requirement on $\Xibm$ simulation as in data.
To test the sensitivity to the position of the points along the decay time axis (in Fig.~\ref{fig:CorrYieldRatio}), 
the fit is repeated assuming an exponential distribution with $\tau=1.80$~ps. 
Bias due to the small signal size has been studied using pseudoexperiments, and we find a small fit bias in
$\tauOmb/\tauXib$, which pulls the value down by 10\% of the statistical uncertainty.
We correct the data for this bias, and assign half the shift as a systematic uncertainty.
The simulated samples used to determine the relative efficiency are of finite size, and those uncertainties are propagated to the final result.
\begin{table*}[h]
\begin{center}
\caption{\small{Summary of systematic uncertainties in $\delta m$ and the lifetime ratio. When two values are indicated, the first is a correction, and
the second is the uncertainty.}}
\begin{tabular}{lcc}
Source & $\delta m$ & $\tauOmb/\tauXib$ \\
           &   (\mevcc)        &                  \\
\hline
Signal shape           &  $\pm$0.3 & $\pm$0.005 \\
Background shape       &  $\pm$0.1 & $\pm$0.009 \\
$\OmegacStar$ shape   & $\pm$0.1  & $\pm$0.003 \\
$X_b\to X_c\Km$ background & $\pm$0.2  & $\pm$0.002 \\
Relative efficiency    &  --   &  $\pm$0.018 \\
Average time in bin    &  --   &  $\pm$0.002 \\
Lifetime fit           &  -- &  $+0.016\pm0.008$ \\
Simulated sample size  &  $-0.38\pm0.28$ & $\pm$0.017 \\
Momentum scale         &  $\pm$0.1 &    --  \\
$\Xibm$ lifetime       &   --   & $\pm$0.004 \\
\hline
Total systematic             &    $-0.4\pm0.5$      &  $+0.016\pm0.029$     \\
\hline
Total statistical             &    $\pm$3.2     &   $\pm$0.16     \\
\end{tabular}
\label{tab:SummaryOfShifts}
\end{center}
\end{table*}

For the $\delta m$ measurement, the fitted value of $\delta m_{\rm meas} - \delta m_{\rm true}$ in simulation is \mbox{$-0.38\pm0.28$\mevcc}.
We apply this value as a correction, and assign the 0.28\mevcc as a systematic uncertainty.
The momentum scale has a fractional uncertainty of $\pm0.0003$~\cite{LHCb-PAPER-2013-011}. Its effect is evaluated by shifting all
momentum components of the final-state particles by this amount in simulated decays, and comparing to the case when no shift is applied.
Lastly, the uncertainty in the $\Xibm$ lifetime enters weakly into the lifetime ratio (see Eq.~\ref{eq:ltratio}), and
is also included as a source of uncertainty.
All sources of systematic uncertainty are added in quadrature to obtain the corrections and systematic uncertainties
of $-0.4\pm0.5$\mevcc on $\delta m$ and $+0.016\pm0.029$ on $\tauOmb/\tauXib$.
\section{Summary}
In summary, a 3.0\invfb $pp$ collision data sample is used to reconstruct 
a sample of $63\pm9$ $\Omegab\to\Omegac\pim$, $\Omegac\to p\Km\Km\pip$ decays. 
This is the first observation of these $\Omegab$ and $\Omegac$ decay modes, with well over
5$\sigma$ significance. Using these signals, the mass difference and mass are measured to be
\begin{align*}
\mOmb-\mXib &= 247.3\pm3.2\pm0.5\mevcc, \\
\mOmb &= 6045.1\pm3.2\pm0.5\pm0.6\mevcc,
\end{align*}
\noindent where the uncertainties are statistical, systematic, and from knowledge of the $\Xibm$ mass~\cite{LHCb-PAPER-2014-048} ($\mOmb$ only).
The measured $\Omegab$ mass is consistent with previous measurements from LHCb,
\mbox{$6046.0\pm2.2\pm0.5$}\mevcc~\cite{LHCb-PAPER-2012-048}, and CDF, \mbox{$6047.5\pm3.8\pm0.6$}\mevcc~\cite{Aaltonen:2014wfa},
but inconsistent with the value of $6165\pm10\pm13$\mevcc obtained by the D0 experiment~\cite{Abazov:2008qm}. An average of
the two LHCb measurements yields $\mOmb = 6045.7\pm1.9\mevcc$, where the momentum scale uncertainty is taken as 100\% correlated,
and the rest of the uncertainties are uncorrelated.

The lifetime ratio and absolute lifetime of the $\Omegab$ baryon are also measured to be
\begin{align*}
\frac{\tauOmb}{\tauXib} &= 1.11\pm0.16\pm0.03, \\
\tauOmb &= 1.78\pm0.26\pm0.05\pm0.06~{\rm ps},
\end{align*}
\noindent using $\tauXib=1.599\pm0.041\pm0.022$~ps~\cite{LHCb-PAPER-2014-048}. The first uncertainty in each
case is statistical. The second uncertainty on $\tauOmb/\tauXib$ is the total systematic uncertainty,
as given in Table~\ref{tab:SummaryOfShifts}. For $\tauOmb$, the second uncertainty is from all sources in
Table~\ref{tab:SummaryOfShifts} except the $\Xibm$ lifetime, and the third uncertainty stems from the uncertainty in 
the $\Xibm$ lifetime. The lifetime is consistent with the previous measurements of
$\tauOmb=1.54^{+0.26}_{-0.21}\pm0.05$~ps~\cite{LHCb-PAPER-2014-010} and
$\tauOmb=1.66^{+0.53}_{-0.40}$~ps~\cite{Aaltonen:2014wfa} by the LHCb and CDF collaborations, respectively. 
The average of the LHCb measurements, assuming no correlation among the uncertainties, yields an $\Omegab$ lifetime 
of $1.66^{+0.19}_{-0.18}$~ps. These measurements improve  
our knowledge of the mass and the lifetime of the $\Omegab$ baryon.
Due to the similarity of the signal and calibration modes,
this pair of decay modes is very promising for future studies of the $\Omegab$ baryon.


\section*{Acknowledgements}
 
\noindent We express our gratitude to our colleagues in the CERN
accelerator departments for the excellent performance of the LHC. We
thank the technical and administrative staff at the LHCb
institutes. We acknowledge support from CERN and from the national
agencies: CAPES, CNPq, FAPERJ and FINEP (Brazil); NSFC (China);
CNRS/IN2P3 (France); BMBF, DFG and MPG (Germany); INFN (Italy); 
FOM and NWO (The Netherlands); MNiSW and NCN (Poland); MEN/IFA (Romania); 
MinES and FANO (Russia); MinECo (Spain); SNSF and SER (Switzerland); 
NASU (Ukraine); STFC (United Kingdom); NSF (USA).
We acknowledge the computing resources that are provided by CERN, IN2P3 (France), KIT and DESY (Germany), INFN (Italy), SURF (The Netherlands), PIC (Spain), GridPP (United Kingdom), RRCKI and Yandex LLC (Russia), CSCS (Switzerland), IFIN-HH (Romania), CBPF (Brazil), PL-GRID (Poland) and OSC (USA). We are indebted to the communities behind the multiple open 
source software packages on which we depend.
Individual groups or members have received support from AvH Foundation (Germany),
EPLANET, Marie Sk\l{}odowska-Curie Actions and ERC (European Union), 
Conseil G\'{e}n\'{e}ral de Haute-Savoie, Labex ENIGMASS and OCEVU, 
R\'{e}gion Auvergne (France), RFBR and Yandex LLC (Russia), GVA, XuntaGal and GENCAT (Spain), Herchel Smith Fund, The Royal Society, Royal Commission for the Exhibition of 1851 and the Leverhulme Trust (United Kingdom).


\clearpage

\ifx\mcitethebibliography\mciteundefinedmacro
\PackageError{LHCb.bst}{mciteplus.sty has not been loaded}
{This bibstyle requires the use of the mciteplus package.}\fi
\providecommand{\href}[2]{#2}


\newpage

\centerline{\large\bf LHCb collaboration}
\begin{flushleft}
\small
R.~Aaij$^{39}$, 
C.~Abell\'{a}n~Beteta$^{41}$, 
B.~Adeva$^{38}$, 
M.~Adinolfi$^{47}$, 
Z.~Ajaltouni$^{5}$, 
S.~Akar$^{6}$, 
J.~Albrecht$^{10}$, 
F.~Alessio$^{39}$, 
M.~Alexander$^{52}$, 
S.~Ali$^{42}$
% ...

{\footnotesize \it
% ...
$ ^{5}$Clermont Universit\'{e}, Universit\'{e} Blaise Pascal, CNRS/IN2P3, LPC, Clermont-Ferrand, France\\
$ ^{6}$CPPM, Aix-Marseille Universit\'{e}, CNRS/IN2P3, Marseille, France\\
% ...
$ ^{10}$Fakult\"{a}t Physik, Technische Universit\"{a}t Dortmund, Dortmund, Germany\\
% ...
$ ^{38}$Universidad de Santiago de Compostela, Santiago de Compostela, Spain\\
$ ^{39}$European Organization for Nuclear Research (CERN), Geneva, Switzerland\\
% ...
$ ^{41}$Physik-Institut, Universit\"{a}t Z\"{u}rich, Z\"{u}rich, Switzerland\\
$ ^{42}$Nikhef National Institute for Subatomic Physics, Amsterdam, The Netherlands\\
% ...
$ ^{47}$H.H. Wills Physics Laboratory, University of Bristol, Bristol, United Kingdom\\
% ...
$ ^{52}$School of Physics and Astronomy, University of Glasgow, Glasgow, United Kingdom\\
% ...
}
\end{flushleft}

\end{document}